\documentclass[11pt]{article}

\usepackage{jheppub} 
 \usepackage{graphicx}                     
 \usepackage{epstopdf}                     
\usepackage[T1]{fontenc} 

\graphicspath{{/}}  

\title{\boldmath One loop amplitude from null string}

\author[a,c]{Ming Yu,}
\author[a,c]{Chi Zhang}
\author[a,b]{and Yao-Zhong Zhang}

\affiliation[a]{CAS Key Laboratory of Theoretical Physics, Institute of Theoretical Physics, Chinese Academy of Sciences,  Beijing 100190, China}
\affiliation[b]{School of Mathematics and Physics, The University of Queensland,  Brisbane Qld 4072, Australia}
\affiliation[c]{School of Physical Sciences, University of Chinese Academy of Sciences, Beijing 100049, China}

\emailAdd{yum@itp.ac.cn}
\emailAdd{zhangchi@itp.ac.cn}
\emailAdd{yzz@maths.uq.edu.au}

\abstract{We generalize the CHY formalism to one-loop level, based on the framework of the null string theory. The null string, a tensionless string theory, produces the same results as the ones from the chiral ambitwistor string theory, with the latter believed to  give a string interpretation of the CHY formalism. A key feature of our formalism is the interpretation of the modular parameters. We find that the $S$ modular transformation invariance of the ordinary string theory does not survive in the case of the null string theory. Treating the integration over the modular parameters this way enable us to derive the n-gons scattering amplitude in field theory, thus proving the n-gons conjecture.}

\newcommand{\dif}{\mathrm{d}} 
\newcommand{\mi}{\mathrm{i}}  

\begin{document}
\maketitle
\flushbottom

\section{Introduction}

When one compares the calculations of Feynman diagrams in field theory with those of the corresponding correlation functions in string theory, one would realize that a conceptual simplification occurs in the latter which reduces the number of diagrams involved. Of course, to make the comparison manifest, one has to take the $\alpha' \rightarrow 0$ limit  in the string theory.
On the other hand, in a series of remarkable papers
\cite{cachazo2014scattering,cachazo2013scattering,cachazo2014scatteringk}, Cachazo, He and Yuan formulated the tree-level scattering amplitudes of massless particles in field theory in terms of 2d chiral conformal field theories.
The main ingredient of the CHY formulation is the on-shell scattering equation,
\[
\sum_{j\neq i} \frac{k_{i}\cdot k_{j}}{z_{i}-z_{j}}=0 \text{ .}
\]
It now appears that this formalism is a consequence of the ambitwistor string theory \cite{mason2013ambitwistor}. This is rather surprising since the ambitwistor string theory is a tensionless (or null) string theory which corresponds to the limit opposite to the usual string-particle equivalence one $\alpha’\rightarrow 0$. However, cf. \cite{casali2016null} for an explanation.
Keeping this underlying string theory in mind, it is natural to ask the question of how to generalize the tree level formulation to one- and multi-loop cases.

Ideas in such direction have recently been explored by several authors \cite{adamo2013ambitwistor,geyer2015one,geyer2015loop,casali2014infrared,he2015one,cachazo2015one}, see also \cite{Baadsgaard2015Integration,Baadsgaard2015New} for a different approach. Two new elements are found to get involved in the scattering equatons.
One is the loop momentum which is the zero mode of the spacetime field, and the other one is the elliptic function $\theta_{1}$. It turns out that one does not need to solve the scattering equations on the torus involving $\theta_{1}$.
Rather, via integration by parts the problem is reduced to solving some equations on Riemann sphere obtained by taking the  $\tau_{2}\to\infty$ limit. These lead to the following \emph{off-shell} scattering equations
\[
\frac{\ell\cdot k_{i}}{z_{i}} + \sum_{j\neq i} \frac{k_{i}\cdot k_{j}}{z_{i}-z_{j}}=0 \text{ .}
\]

Unlike the on-shell external momenta $k_i, \ \ k_i^2=0$, the off-shell loop momentum $\ell, \ \ell^2\neq 0$ is an integration variable. The off-shell scattering equation is one of the main ingredients of the ambitwistor string theory.
In the field theory limit,  the off-shell tree diagrams can be obtained from the relevant loop diagrams  by cutting off any internal lines in the loop, leaving two punctures on the sphere.
Conversely,  starting from such tree diagrams,  one can recover the loop amplitude by sewing together the two punctures represented by the off-shell momenta. The off-shell scattering equations just tell one how to determine the structure of such tree diagrams,
which perhaps explains the usage of the terminology  ``off-shell''.

However, there are many subtle issues, e.g. modular invariance, which remain to be addressed in the generalization of the CHY formalism to higher genus Riemann surfaces.
In this paper, we present our approach for extending the CHY formalism to the calculation of the one-loop amplitude without the need of worrying about the modular invariance.
From the string theory point of view, the scattering amplitude is just the vacuum expectation value of physical vertex operator insertions. The difference between tree-level and loop-level amplitudes lies in the difference of their corresponding vacuum states.
One might expect that the loop vacuum state inherits the modular invariance originated from the torus topology of the string world-sheet. However, in this paper we shall argue that the modular invariance of the ordinary string theory would not survive in the case of the ambitwistor string theory.
To see this let us start with the null string theory, which is equivalent to the chiral ambitwistor string formulation under certain conditions.  The advantage of the null string theory is that it can  be seen as the ultra-relativistic limit of the string theory, in which the spacetime coordinates are expanded linearly in worldsheet time direction $\tau$ (not to be confused with the modular parameter $\tau$ on the torus) like $e^{2\pi n(\mi\sigma+\tau)}\sim e^{2\pi n(\mi\sigma)}(1+2\pi n\tau)$.
The zero modes are already linear in time direction even before the ultra-relativistic limit is taken, so their contribution to the partition function on the torus remains intact, $ Z_0=\int e^{\mi2\pi(\tau-\bar{\tau})\ell^2} d^d\ell \sim (\tau-\bar{\tau})^{-d/2}$, here $\tau$ is the modular parameter of the torus.
 However the non-zero modes  coming from both left and right moving parts combine to form a chiral theory after the limit is taken, so their contributions become $Z_{\text{osc}}\sim(\prod_{n=1}^{\infty}(1-q^n))^{-2d}$, with $q=\exp (2\pi \mi\tau )$. Taking all these into account, we expect that the partition function looks like
\begin{align*}
Z(\tau, \bar{\tau}) &=Z_0 \, Z_{\text{osc}}  \\
&\sim (\tau-\bar{\tau})^{-d/2}\left(\prod_{n=1}^{\infty}(1-q^n)\right)^{-2d} \text{ .}
\end{align*}

Since $Z_{\text{osc}}$ is holomorphic in $\tau$ while $Z_0$ is not, the combined partition function $Z=Z_0 Z_{\text{osc}}$ is invariant under the $T$ modular transformation, $\tau\rightarrow \tau+1$, but not under the $S$ modular transformation $\tau\rightarrow -1/\tau$. Of course the full partition function should also have contribution from ghosts, but that would only change the chiral part at most. Without $S$ modular invariance, the integration in modular space in our formalism is carried out over the half infinite cylinder $\tau_1 \in [-1/2,1/2]$, $\tau_2 \in (0, \infty)$.\footnote{The author in \cite{Ohmori2015Worldsheet} showed that the moduli space for the ambitwistor string is cotangent to the moduli space of Riemann surfaces, but did not specify the moduli integration limits.} It is clear from this simple example, that upon integration over $\tau_1$, we project out all the (non-zero mode) oscillator contributions and come back to point particle theory, where  $\tau_2$  plays the role of the Schwinger parameter. Consequently we can choose to first integrate either the loop momentum $\ell$ or the Schwinger parameter $\tau_2$, exactly as what one does in point particle field theory. The integration over $\tau_1$ just plays the role of projecting to the oscillator zero modes. We can further assume that there is a one to one correspondence between higher loop Feynman diagrams in field theory and correlators in null string theory on  Riemann surfaces. On the field theory side, we assume three point interactions and consider general multi-loop Feynman diagrams.  Let $n$ be the number of external momenta, $m$ the number of propagators  and $g$ the number of loop momenta to be integrated. Then we find the following simple relation
\begin{align*}
m&=3(g-1)+n, \qquad {\rm for} ~~ g >1, \\
m&=g+n-1, \qquad\quad\, {\rm for} ~~ g =1, \\
m&=g+n-3, \qquad \quad \, {\rm for} ~~ g=0 \text{ .}
\end{align*}
It follows that $m$ coincides exactly with the dimension of the modular space of genus $g$ Riemann surface with $n$ punctures.
With each propagator we can associate a Schwinger parameter. Intuitively the Schwinger parameter would become the imaginary part of the modular parameter in our null string theory.  Although it is known that Schwinger parameters are the origin of modular space in ordinary string theory \cite{Bern1992loop}, it is remarkable that there exists also such an exact correspondence in the case of the ambitwistor string theory.

\section{A quick review of null string}

Null string, or tensionless string, is a class of strings with the action
\begin{equation}
S=\int \dif^{2}\sigma \,V^{\alpha}V^{\beta}\partial_{\alpha}Y \cdot \partial_{\beta}Y \text{ ,}\label{z.0.1}
\end{equation}
which is known as the Lindstr\"{o}m-Sundborg-Theodoris \cite{Karlhede1986wb,lindstrom1991zero} (LST) tensionless string action.
In this paper, we choose the so-called HSZ gauge in Siegel's paper \cite{siegel2015amplitudes,Hohm2014Doubled,siegel2017Chiral}. Then the action becomes
\begin{equation}
S=\int \dif z\dif\bar{z}\, (\bar{\partial}Y)^{2} \text{ ,} \label{z.0.2}
\end{equation}
where we have dropped the Lorentz index for convenience and adopted the complex coordinates. The equation of motion reads
\[
\bar{\partial}^{2}Y =0 \text{ ,}
\]
and its solution on the cylinder is
\[
Y(z,\bar{z})= X(z)+(\bar{z}-z)P(z) \text{ .}
\]
There are two different vacua for the tensionless string. One is the high-spin string vacuum which can be derived from the tensionless limit of the ordinary bosonic string \cite{lizzi1986quantization,gamboa1990null,Isberg:1993av}. The other \cite{Gamboa1989quantum} conforms with the ambitwistor string \cite{casali2016null}. See also \cite{Huang2016Factorization}.
For the second vacuum, the behavior of the $XP$-system is the same as the $\gamma,\beta$ fields with conformal weight (0,0) and (1,0) respectively. We set
$z=\sigma+\mi\tau$, where $\sigma \in [-1/2,1/2]$ and $\tau\in\mathbb{R}$  are the coordinates of the cylinder. On the cylinder, we have the following mode expansion,
\begin{align}
X(z)&=\sum_{n\in \mathbb{Z}}x_n e^{-2\pi\mi nz}\text{ ,}  \\
P(z)&= 2\pi\mi\sum_{n\in \mathbb{Z}}p_n e^{-2\pi\mi nz} \text{ ,} \\
[p_n,x_m]&=\delta_{n+m,0}\text{ ,}
\end{align}
where we have introduced the factor $2\pi \mi$ in $P(z)$ for later convenience. The 2-point function on the cylinder can be calculated as
\begin{equation}
\langle0|P(z)X(z')|0\rangle =2\pi\mi \sum_{n\geq 0}\left(\frac{e^{2\pi \mi z'}}{e^{2\pi\mi z}}\right)^n=\frac{2\pi \mi e^{2\pi\mi z}}{e^{2\pi \mi z}-e^{2\pi \mi z'}}=\pi \left(\cot \pi(z-z')+\mi\right).
\end{equation}
We will ignore the regular term in $\langle PX\rangle$ since such a term is irrelevant for our purpose. In such case  we have  $\langle0|P(z)X(z')|0\rangle =-\langle0|X(z)P(z')|0\rangle $. Then we obtain the correlator of $Y$'s, namely the Green function on the infinite cylinder
\begin{equation}
G(z-z',\bar{z}-\bar{z}')\equiv \langle Y(z,\bar{z})Y(z',\bar{z}')\rangle =
[(\bar{z}-\bar{z}')-(z-z')]\pi\cot\pi(z-z')  \label{z.0.5}\text{ ,}
\end{equation}%
which obviously satisfies the Green equation
\begin{equation}
\bar{\partial}^{2}G(z,z')=2\pi\sum_{n\in\mathbb{Z}}\delta^{2}(z-z'+n).
\end{equation}

This argument verifies that $X(z)$ and $P(z)$ are pair of bosonic $\gamma\beta$ system on the cylinder, with conformal
weight $0,\ \ 1$ respectively. The physical states are the same as in the ambitwistor string theory and given by $p^{\mu}_{-1}p^{\nu}_{-1}|0\rangle$, $(p^{\mu}_{-1}x^{\nu}_{-1}-x^{\mu}_{-1}p^{\nu}_{-1})|0\rangle$, and $p_{-1}^{\mu}p_{-1}^{\mu}|0\rangle$, where $x^{\mu}_{n}$ and $p_{n}^{\mu}$ are respectively the operator coefficients of the mode expansion of $X(z)$ and $P(z)$. Accordingly, we shall see that the one-loop amplitude, at least in our simple example, can be calculated by introducing propagators on the torus.

\section{Path integral evaluation}

In path integral formalism, the scattering amplitude is just one special case of the generating functional
\begin{align}
Z[J] &=\left\langle \exp \left( \mi\int \dif^{2}\sigma\, J(\sigma )\cdot Y(\sigma
)\right) \right\rangle   \nonumber \\
&=\int [\dif Y]\exp \left( -\int \dif^{2}\sigma\, Y(\sigma )DY(\sigma )+\mi J(\sigma
)\cdot Y(\sigma )\right) \text{ ,}  \label{z.1.1}
\end{align}%
where $D$ is certain kind of differential operator. For example, for the bosonic string $D$ is the Laplacian $\nabla ^{2}$. However, for the action we are considering, the operator $D$ is $\bar{\partial}^{2}$ in complex coordinates.
No matter what $D$ is, the path integral is Gaussian. So the result of integration is
\begin{equation}
Z[J]=C\delta ^{d}(J^{0})\left( \det\nolimits^{\prime }\frac{D}{2\pi }\right)
^{-d/2}\exp \left( -\frac{1}{2}\int \dif^{2}\sigma \int \dif^{2}\sigma ^{\prime
}\,J(\sigma )\cdot J(\sigma ^{\prime })G(\sigma ,\sigma ^{\prime })\right)
\text{ .}  \label{z.1.2}
\end{equation}%
Here $d$ is the spacetime dimension of the theory we are considering, $\delta
^{d}(J^{0})$ come from the zero modes of the operator $D$, the primed determinant means that the zero eigenvalues are excluded, $G(\sigma ,\sigma
^{\prime })$ is the Green's function of the operator $D$, and $C$ is an overall
constant. Once the base manifold is specified, what we need to calculate is the determinant
and the Green's function of $D$ on this manifold. For our purpose, the manifold
is a torus and the operator $D$ is $\bar{\partial}^{2}$. It is convenient to rewrite Eq.(\ref{z.1.2}) as
\begin{equation}
Z[J]=C\delta ^{d}(J^{0})\left( \det\nolimits^{\prime }\frac{D}{2\pi }\right)
^{-d/2}\exp \left( -\frac{1}{2}\sum_{i\neq j}k_{i}\cdot k_{j}G(\sigma
_{i},\sigma _{j})\right)\mathcal{I}(\tau,\{z_{i}\}) \text{ ,}  \label{z.1.3}
\end{equation}%
where $\exp(\sum k_{i}\cdot k_{j} G)$ comes from the contraction $\langle :e^{\mi k_{1}\cdot Y(z_{1})}:\cdots :e^{\mi k_{n}\cdot Y(z_{n})}:\rangle$. The reason that we isolate such terms is that they are associated with the $\bar{z}_i$ coordinates,  integration upon which will yield the scattering equation. The other parts are holomorphic in $z_i$ coordinates, which are evaluated according to the solutions of the scattering equation.  However, instead of integrating out $\bar{z}_i$'s to obtain the scattering equation as Siegel did in \cite{siegel2015amplitudes}, in this paper we choose to deal with such terms in a different way at the one-loop level.

\section{Green function}

The Green's function can be viewed as the inverse of the operator $D$. For our purpose, $D=\bar{\partial}^{2}$, thus the equation satisfied by  the Green's function is%
\begin{equation}
\bar{\partial}^{2}G=2\pi \delta ^{2}(z)\text{ .}  \label{z.2.1}
\end{equation}%
On the infinite cylinder, its solution is given by
\begin{equation}
G(z)=(\bar{z}-z)\pi \cot (\pi z)\text{ .} \label{z.2.3}
\end{equation}%
However, on torus we have to take into account the background charge, which gives rise to the following Green's equation
\begin{equation}
\bar{\partial}^{2}G=2\pi \delta ^{2}(z)-\frac{\pi }{\tau _{2}} . \label{z.2.4}
\end{equation}%
Its solution can be shown to be
\begin{equation}
G(z)=(\bar{\tau}-\tau )\partial _{\tau }\ln \theta _{1}(z|\tau )+(\bar{z}%
-z)\partial _{z}\ln \theta _{1}(z|\tau )+\frac{\pi \mi}{\bar{\tau}-\tau }(%
\bar{z}-z)^{2} + f(\tau,\bar{\tau})\text{ ,}   \label{z.2.5}
\end{equation}%
where%
\begin{equation*}
\theta _{1}(z|\tau ) =-\mi\sum_{n=-\infty }^{\infty
}(-1)^{n}q^{(n-1/2)^{2}/2}e^{2\mi\pi z(n-1/2)} \text{ .}
\end{equation*}%
The function $f(\tau ,\bar{\tau})$ is determined
by the fact that $G(z)\sim \bar{z}/z$ as $z\rightarrow 0$, but its
contribution to the amplitude vanishes due to the momentum conservation, so we do not need to concern ourselves with its exact form. It is straightforward to check the two periods of the Green
function (\ref{z.2.5})%
\begin{equation}
z\rightarrow z+1\text{, \ \ \ \ }z\rightarrow z+\tau \text{,}  \label{z.2.6}
\end{equation}%
In the remainder of this article, we will ignore the term $f(\tau ,\bar{\tau})$ since it has no contribution to the amplitudes concerned. We will give an intuitive derivation of this Green function in the appendix.

\section{Partition function}

Now, let us consider the simplest case $J=0$, that is, there is no vertex operator insertion in the path integral.
The result is also known as the partition function, denoted as $Z(\tau)$. According to Eq. (\ref{z.1.3}), for $d=1$, the partition function is just the determinant of the operator $\bar{\partial}^{2}$, so we have
\begin{equation}
Z(\tau ) = \sqrt{A}\prod_{n}\left( \frac{2\pi }{\lambda _{n}}\right) ^{1/2}%
\text{ ,}  \label{z.3.1}
\end{equation}%
where $A$, coming from the zero mode integration, is the area of the torus and $\lambda _{n}$ are the nonzero eigenvalues of the operator $\bar{\partial}^{2}$ \cite{francesco2012conformal}.

The eigen equation of $\bar{\partial}^{2}$ is%
\begin{equation}
\bar{\partial}^{2}f=\lambda _{n}f\text{ ,}  \label{z.3.2}
\end{equation}%
so we obtain the eigen function%
\begin{align*}
f &=\exp \left( 2m\pi \mi\frac{z-\bar{z}}{\tau -\bar{\tau}}\right) \exp
\left( 2n\pi \mi\left( \frac{z}{\tau }-\frac{\bar{z}}{\bar{\tau}}\right) \frac{%
1}{1/\tau -1/\bar{\tau}}\right)  \\
&=\exp \left( 2\pi \mi z\frac{\left( m-n\bar{\tau}\right) }{\tau -\bar{\tau}}%
\right) \exp \left( 2\pi \mi\bar{z}\frac{-m+n\tau }{\tau -\bar{\tau}}\right)
\end{align*}%
and eigenvalues%
\begin{equation}
\lambda _{n}=\left( 2\pi \mi\frac{m-n\tau }{\tau -\bar{\tau}}\right) ^{2}\text{
.}  \label{z.3.3}
\end{equation}%
Then the partition function becomes%
\begin{align}
Z(\tau ) &=\sqrt{A}\prod_{m,n}\left( 2\pi \left( \mi\frac{m-n\tau }{%
\tau -\bar{\tau}}\right) ^{2}\right) ^{-1/2}  \nonumber \\
&=\sqrt{A}\prod_{m,n}\frac{\sqrt{2}\tau _{2}}{\sqrt{\pi }}\frac{1}{m+n\tau }%
\text{ .}  \label{z.3.4}
\end{align}%
We use the $\zeta $-function regularization technique and define%
\begin{equation}
G(s)\equiv \sum\nolimits^{\prime }\left( \frac{\sqrt{2}\tau _{2}}{\sqrt{\pi }}%
\frac{1}{m+n\tau }\right) ^{s}\text{ ,}  \label{z.3.5}
\end{equation}%
where the primed sum means that the zero eigenvalues are excluded. The partition function is then formally expressed as
\begin{equation}
Z(\tau )=\sqrt{A}\exp G^{\prime }(0)\text{ .}  \label{z.3.6}
\end{equation}

Now we compute $G(s)$. It is convenient to express it as%
\begin{align}
\left( \frac{\sqrt{\pi }}{\sqrt{2}\tau _{2}}\right) ^{s}G(s) &=\sum
\nolimits^{\prime}\left( \frac{1}{m+n\tau }\right) ^{s}  \nonumber \\
&=(1+(-1)^{-s})\zeta (s)+\sum_{n}\nolimits^{\prime}\left( \sum_{m}\frac{1}{%
\left( m+n\tau \right) ^{s}}\right)   \label{z.3.7}
\end{align}%
where $\zeta(s)$ is the Riemann $\zeta$-function. The second term on the r.h.s of Eq. (\ref{z.3.7}) is a periodic function of $n\tau$ with unit period and may therefore be Fourier expanded
\begin{align}
\quad\sum_{m}\frac{1}{\left( m+n\tau \right) ^{s}}
&=\sum_{p}e^{2\mi\pi pn\tau _{1}}\int_{0}^{1}\dif y\,e^{-2\pi \mi py}\sum_{m}\frac{1}{%
\left( m+y+\mi n\tau _{2}\right) ^{s}} \nonumber \\
&=\sum_{p}e^{2\mi\pi pn\tau _{1}}\int_{-\infty }^{\infty }\dif y\, e^{-2\pi \mi py}%
\frac{1}{\left( y+\mi n\tau _{2}\right) ^{s}} \label{z.3.8}
\end{align}%

Now, we consider the integral%
\[
\sum_{n\neq 0}\sum_{p}\int_{-\infty }^{\infty }\dif y \,e^{2\mi\pi p(n\tau _{1}-y)}%
\frac{1}{\left( y+\mi n\tau _{2}\right) ^{s}} \text{ .}
\]%
This sum can be decomposed into five parts: (i) $p>0$ and $n>0$, (ii) $p>0$
and $n<0$, (iii) $p<0$ and $n>0$, (iv) $p<0$ and $n<0$, (v) $p=0$. When $p$ is
less than zero, we can close the contour in the upper half plane. Then the part with $n>0$ has
no contribution since the branch point is at $-\mi n\tau _{2}$. Similarly, when $p>0$, we close the contour in the lower half plane and only the part with $n>0$ contributes. Thus, apart from the case $p=0$, we just need to consider the integration%
\[
\sum_{n<0}\sum_{p<0}\int_{-\infty }^{\infty }\dif y\,e^{-2\mi\pi py}\frac{1}{\left(
y+\mi n\tau _{2}\right) ^{s}}+\sum_{n>0}\sum_{p>0}\int_{-\infty }^{\infty
}\dif y\, e^{-2\mi\pi py}\frac{1}{\left( y+\mi n\tau _{2}\right) ^{s}}\text{ .}
\]%
\begin{figure}[!t]
  \centering
  \includegraphics[width=3.5in]{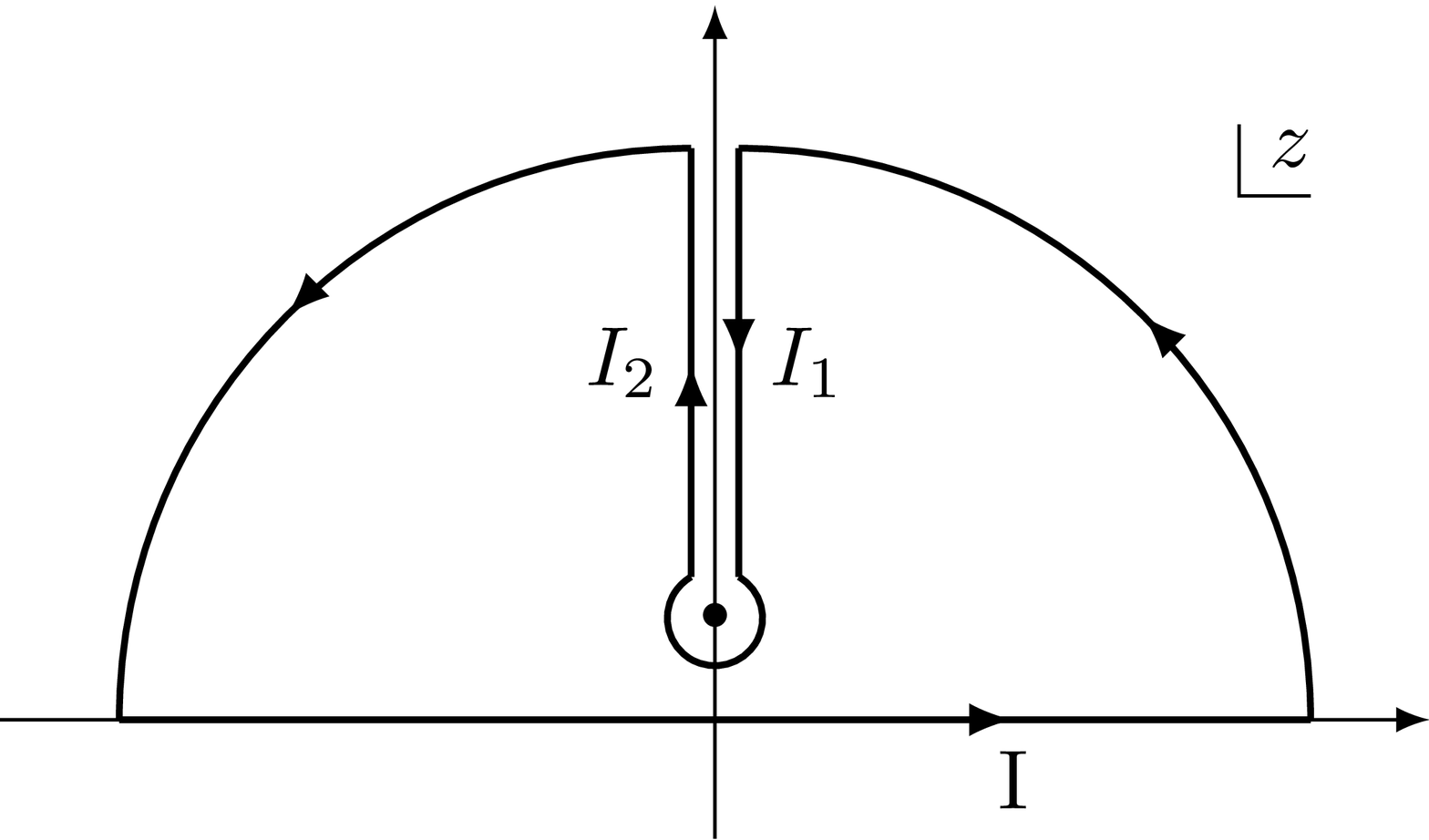}
   \caption{}
 \label{fig:1}
\end{figure}
We choose the integration contour for the first term as  depicted in the Fig \ref{fig:1}. Applying the Cauchy integral formula%
\begin{equation}
\int_{-\infty }^{\infty }\dif y\, e^{-2\mi\pi py}\frac{1}{\left( y+\mi n\tau
_{2}\right) ^{s}} =I =-I_{1}-I_{2}   \qquad(\text{for }p<0 \text{ and } n<0 ) \text{ ,}\label{z.3.9}
\end{equation}%
where
\begin{align*}
I_{1}&=\mi\int_{\infty }^{-n\tau _{2}}\dif x\, e^{2\pi px}\frac{e^{-\mi\pi s/2}}{(x+n\tau
_{2})^{s}} =-\mi e^{-\mi\pi s/2}e^{-2np\pi \tau _{2}}(-p)^{s-1}(2\pi
)^{s-1}\Gamma (1-s) \\
I_{2}&=\mi\int_{-n\tau _{2}}^{\infty }\dif x\, e^{2\pi px}\frac{e^{\mi 3\pi s/2}}{%
(x+n\tau _{2})^{s}}=\mi e^{\mi 3\pi s/2}e^{-2np\pi \tau _{2}}(-p)^{s-1}(2\pi
)^{s-1}\Gamma (1-s) \text{ .}
\end{align*}
So we get%
\[
I=-e^{\mi\pi(1-s)/2}(e^{2\mi\pi s}-1)e^{-2np\pi \tau _{2}}(-2p\pi
)^{s-1}\Gamma (1-s)\text{ .}
\]%
Similarly  for $n>0$ and $p>0$, we obtain
\begin{equation}
\int_{-\infty }^{\infty }\dif y\, e^{-2\mi\pi py}\frac{1}{\left( y+\mi n\tau
_{2}\right) ^{s}} =-e^{-\mi\pi(1-s)/2}(e^{-2\mi\pi s}-1)e^{-2np\pi \tau
_{2}}(2\pi p )^{s-1}\Gamma (1-s) \text{ .}\label{z.3.10}
\end{equation}
For the case $p=0$, the direct evaluation of the first line of eq.(\ref{z.3.8}) yields,
\begin{align*}
\tilde{I}(\tau_{2},s)&\equiv\int_{0 }^{1 }\dif y \,%
 \sum_{m}\frac{1}{\left( m+y+\mi n\tau _{2}\right) ^{s}} \\
 &= \frac{1}{1-s} \left(\sum_{m}\frac{1}{( m+1+\mi n\tau _{2}) ^{s-1}} -\sum_{m}\frac{1}{( m+\mi n\tau _{2}) ^{s-1}}\right)=0 \text{ .}
\end{align*}
Thus, Eq. (\ref{z.3.8}) yields%
\begin{align*}
&\sum_{n>0}\sum_{p>0}-e^{-\mi\pi(1-s)/2}(e^{-2\mi\pi s}-1)e^{-2np\pi \tau
_{2}}(2\pi p )^{s-1}\Gamma (1-s) e^{2\mi\pi pn\tau _{1}} \\
&\qquad\qquad +\sum_{n<0}\sum_{p<0}-e^{\mi\pi(1-s)/2}(e^{2\mi\pi s}-1)e^{-2np\pi \tau _{2}}(-2p\pi
)^{s-1}\Gamma (1-s)e^{2\mi\pi pn\tau _{1}} \\
&\qquad =\sum_{n>0}\sum_{p>0}\frac{2}{(2\pi )^{1-s}}\left(\cos (\frac{\pi(1- s)}{2})-\cos (\frac{\pi(1+3s)}{2})\right)e^{-2np\pi \tau
_{2}}p^{s-1}\Gamma (1-s)e^{2\mi\pi pn\tau _{1}} \\
&\qquad=2s\sum_{n>0}\sum_{p>0}(p)^{-1}e^{2\mi\pi pn\tau }+O(s^{2}) \\
&\qquad=-2s\ln \prod_{n>0}(1-q^{n})+O(s^{2})\text{ .}
\end{align*}%

Putting them all together we can now write Eq.(\ref{z.3.7}) as
\begin{equation}
G(s)=\left( \frac{\sqrt{\pi }}{\sqrt{2}\tau _{2}}\right) ^{-s}\left[
(1+(-1)^{-s})\zeta (s)-2s\ln \prod_{n>0}(1-q^{n})\right] +O(s^{2})  \text{ .}\label{z.3.11}
\end{equation}%
\newline
We thus obtain the partition function%
\begin{align}
Z(\tau ) &=\sqrt{A}\exp G^{\prime }(0)
=\sqrt{\tau _{2}}(-1)^{1/2}\frac{1}{\sqrt{8\pi }\tau _{2}}%
\prod_{n>0}(1-q^{n})^{-2}  \nonumber\\
&=\frac{\mi}{\sqrt{8\pi \tau _{2}}}\prod_{n>0}(1-q^{n})^{-2} \text{ .} \label{z.3.12}
\end{align}

We now integrate out variable $\tau$ in the expression (\ref{z.3.12}).
To carry out this integration, we shall integrate first
over $\tau _{1}$. Since only the infinite product has dependence on
$\tau _{1}$, this integral simplifies to%
\begin{align}
Z(\tau _{2}) &=\int_{-1/2}^{1/2}\dif\tau _{1}\,Z(\tau )=\frac{\mi}{\sqrt{8\pi \tau _{2}}}\int_{-1/2}^{1/2}\dif\tau
_{1}\,\prod_{n>0}(1-q^{n})^{-2}  \nonumber \\
&=\frac{\mi}{\sqrt{8\pi \tau _{2}}}\int_{-1/2}^{1/2}\dif\tau _{1}\,\left(
\sum_{n=0}^{\infty }p(n)q^{n}\right) ^{2} \text{ ,} \label{z.3.15}
\end{align}%
where $p(n)$ is the partition number of $n$. Since for $|q|<1$, the power series on the r.h.s. of eq.(\ref{z.3.15}) converges, we have,  for any nonzero integer $m$,%
\begin{equation}
\int_{-1/2}^{1/2}\dif\tau _{1}\,q^{m}=\int_{-1/2}^{1/2}\dif \tau _{1}\,e^{2m\pi i\tau
_{1}}e^{-2m\pi \tau _{2}}=0 \text{ .} \label{z.3.16}
\end{equation}%
So we obtain the zero point one-loop amplitude for the null string%
\begin{equation}
A_{\text{one-loop}}=\int_{0}^{\infty}(Z(\tau _{2}))^d\dif\tau_{2}=\int_{0}^{\infty}\dif\tau_{2}\,(\frac{\mi}{\sqrt{8\pi \tau _{2}}})^d\text{ .}  \label{z.3.17}
\end{equation}%

For comparison, we also give the partition function for massless
particles. As mentioned in the beginning of this section, the partition function for this case is just the one-loop vacuum bubble, given by%
\begin{align}
A(\text{vac bubble}) &=\int \frac{\dif^{d}k}{(2\pi )^{d}}\frac{1}{k^{2}}
=\int \frac{\dif^{d}k}{(2\pi )^{d}}\int_{0}^{\infty }\dif l\exp (-k^{2}l)
\nonumber \\
&=\mi\int_{0}^{\infty }\dif l\,(2\pi l)^{-d/2} \text{ ,} \label{z.3.13}
\end{align}%
where we have used the method of Schwinger proper time, and the factor $\mi$ in front of the last line arises form the Wick rotation. This can be considered as the
sum over all particle paths with the topology of a circle, $l$ being the
length of the circle.

We find, apart from some adjustable constant factors, the zero point one-loop amplitude (\ref{z.3.17}) for the null string  is the same as that for the massless particles.
This strongly supports that $\tau _{2}$ is identical to the Schwinger proper time in point particle field theory.

\section{Schwinger form of scattering amplitude}

It was shown in \cite{casali2014infrared} that for supergravity the factor
$\mathcal{I}(\tau,\{z_{i}\})$ in eq.(\ref{z.1.3}) is independent of $z_{i}$ and $\tau$ when $n = 4$, as in conventional string theory. For simplicity, we choose $\mathcal{I}=1$ in this case.

In fact it is not easy to solve the scattering equations on the torus. In the previous section, we have shown that the partition
function can be simplified by first integrating out the $\tau _{1}$ variable. We expect that the same kind of simplification will occur in the computation of amplitudes. Before proceeding, we specify the
domain of $z$-integration to be $\left\vert \operatorname{Re}z\right\vert <1/2$ and $0<
\operatorname{Im}z<\tau _{2}$, so that the integration over $\tau _{1}$ is straightforward. Using the infinite product form of
$\theta_{1}$,
\[
\theta _{1}(z|\tau )=2q^{1/8}\sin (\pi z)\prod_{m=1}^{\infty
}(1-q^{m})(1-e^{2\pi iz}q^{m})(1-e^{-2\pi iz}q^{m})
\]%
we find that $G(z)$ can be written as
\begin{align}
G(z) &=(\bar{\tau}-\tau )\partial _{\tau }\ln \theta _{1}(z|\tau )+(\bar{z}%
-z)\partial _{z}\ln \theta _{1}(z|\tau )+\frac{\pi \mi}{\bar{\tau}-\tau }(\bar{%
z}-z)^{2}  \nonumber \\
&=\frac{\pi \tau _{2}}{2}+(\bar{z}-z)\pi \cot (\pi z)-\frac{\pi }{2\tau _{2}%
}(\bar{z}-z)^{2}+O(q)\text{.}  \label{z.5.1}
\end{align}%
Integrating out $\tau_1$ first, we get the expression for the partial amplitude%
\begin{equation}
A(k_{1},\cdots ,k_{n})=C\int_{0}^{\infty }\dif\tau _{2}(\tau
_{2})^{-d/2}\prod_{\mu =2}^{n}\int \dif z_{\mu }\dif\bar{z}_{\mu }\exp (-
\sum_{i<j}^n k_{i}\cdot k_{j}\tilde{G}_{ij})  \label{z.5.2}
\end{equation}%
where%
\begin{equation}
\tilde{G}_{ij}=\frac{\pi \tau _{2}}{2}+(\bar{z}_{ij}-z_{ij})\pi \cot (\pi
z_{ij})-\frac{\pi }{2\tau _{2}}(\bar{z}_{ij}-z_{ij})^{2}\text{ ,}
\label{z.5.3}
\end{equation}%
$C$ is an overall constant, $z_{ij}\equiv z_i-z_j$ and $n=4$ for the present example. Note that the first term on the r.h.s. of the eq.(\ref{z.5.3})
actually has no contribution to the amplitude since $\sum_{ij}k_{i}\cdot
k_{j}=0$.

\begin{figure}[!t]
  \centering
  \includegraphics[width=5.0in]{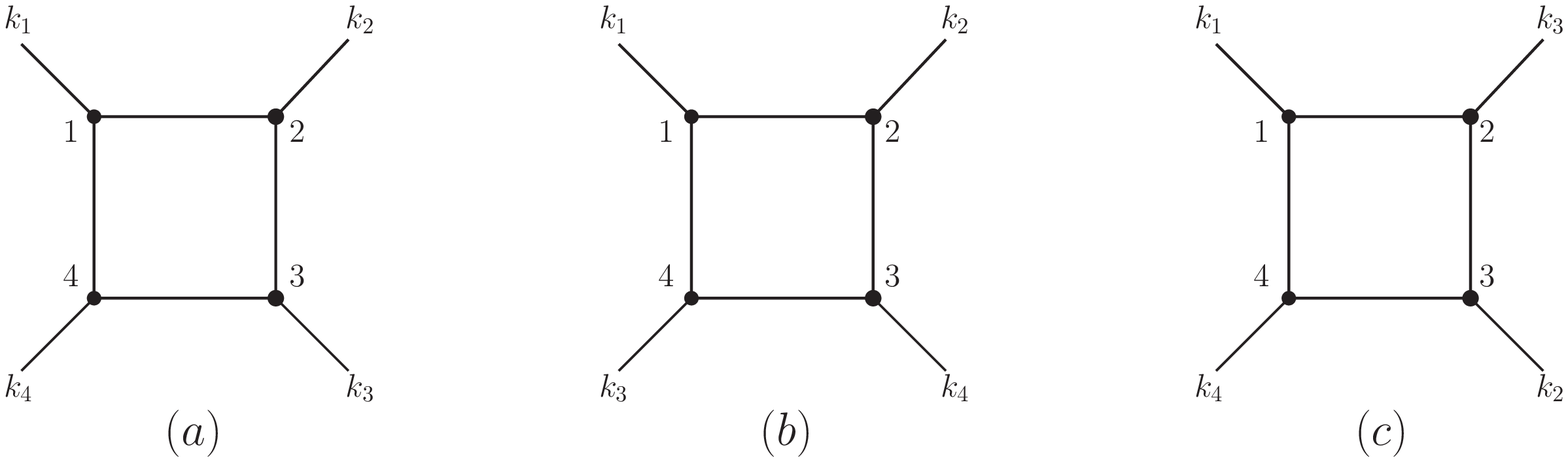}
   \caption{}
 \label{fig:2}
\end{figure}

We can set $z_{1}=0$ by making use of the two conformal Killing vectors on the torus. Then there are three
possible ways in dividing the integration region up to topological equivalence: (i) $0=%
\operatorname{Im}z_{1}<\operatorname{Im}z_{2}<\operatorname{Im}z_{3}<\operatorname{Im}z_{4}<\tau _{2}$, (ii)
$0=\operatorname{Im}z_{1}<\operatorname{Im}z_{2}<\operatorname{Im}z_{4}<\operatorname{Im}z_{3}<\tau _{2}$,
and (iii) $0=\operatorname{Im}z_{1}<\operatorname{Im}z_{4}<\operatorname{Im}z_{2}<\operatorname{Im}%
z_{3}<\tau _{2}$. (As a matter of fact, there are six ways to split the integration region, but the other three cases give the same contributions as those three mentioned above, as will become clear in what follows.) Since the difference
among the cases (i), (ii) and (iii) is just the interchange of the momentum indices, it is sufficient to consider only the case (i). Note that, $\tilde{G}_{ij}$ is regular at $z_{i}=z_{j}$,
since it is suppressed by  $\operatorname{Im}z_{i}-\operatorname{Im}z_{j}$, so that the multi-integral in eq.(\ref{z.5.2}) is well defined.
We will show that Eq. (\ref{z.5.2}) give the
amplitude for the Feynman diagrams in Fig.\ref{fig:2}. More precisely, the case (i) corresponds to figure (a), the case (ii)
to figure (b), and the case (iii)  to figure (c). In what
follows, we will focus on the case (i).

Before doing the integration in Eq. (\ref{z.5.2}), let us review the
Schwinger parametrization in field theory. In terms of Schwinger parameters $s_i$ ($i=1,2,3,4$) the amplitude for massless particles represented by  figure (a) can be written as
\begin{align}
&\int \frac{\dif^{d}p}{(2\pi )^{d}}\frac{1}{%
p^{2}(p+k_{1})^{2}(p+k_{1}+k_{2})^{2}(p-k_{4})^{2}}  \nonumber \\
&\qquad=\int \frac{\dif^{d}p}{(2\pi )^{d}}\int_{0}^{\infty }\dif s_{1}\cdots
\int_{0}^{\infty }\dif s_{4}  \nonumber \\
&\qquad\quad\exp \left[
-s_{4}p^{2}-s_{1}(p+k_{1})^{2}-s_{2}(p+k_{1}+k_{2})^{2}-s_{3}(p-k_{4})^{2}%
\right]   \nonumber \\
&\qquad=C_{1}\prod_{i=1}^{4}\int_{0}^{1}\dif \alpha _{i} \,\delta\left(\sum_{i=1}^{4}\alpha_{i}-1\right)\int_{0}^{\infty
}\dif \tau _{2}(\tau _{2})^{3-d/2}\exp \left[ \tau _{2}(\ell ^{2}-2\alpha
_{2}k_{1}\cdot k_{2})\right] \text{ ,}  \label{z.5.4}
\end{align}%
with%
\begin{align}
\ell  &=k_{1}(\alpha _{1}+\alpha _{2}+\alpha _{3})+k_{2}(\alpha _{2}+\alpha
_{3})+k_{3}\alpha _{3}\text{ ,}  \label{z.5.5} \\
\tau _{2} &=s_{1}+s_{2}+s_{3}+s_{4}\text{ ,}  \label{z.5.6} \\
\alpha _{i}  &=\frac{s_{i}}{\tau _{2}}\text{ ,}\qquad\quad i=1,2,3,4 \text{ ,}\label{z.5.7}\\
k_i^2 &=0\text{ ,}\qquad\quad i=1,2,3,4 \text{ .}\nonumber
\end{align}%
Here $C_{1}$ is the integration constant, and we have eliminated $k_{4}$ by the
conservation of momentum. In this formalism, the geometric meaning of each
parameter is clear. $\tau _{2}$ is the proper time of the whole loop, and $\alpha _{i}\tau _{2}$ is the proper time of the world line connecting the point $i$ and the point $i+1$. We use the same parameter $\tau _{2}$ as what is used in
the string theory because they play the same role. In what follows  we will show the
connection between $\alpha _{i}$ and coordinates $z_{i}$ of the vertex operators, or more specifically, the connection between $\alpha _{i}$
and $\operatorname{Im}z_{i}$.

We now turn to the integration in Eq. (\ref{z.5.2}). For our purpose, we should integrate over $\operatorname{Re}z_{i}$ first. It is convenient to define $\sigma _{i}=\exp (2\pi \mi z_{i})$.
We denote the phase argument of $\sigma _{i}$
as $\phi _{i}$, $\operatorname{Re}z_{i}$ as $x_{i}$, and $\operatorname{Im}z_{i}$ as $y_{i}$, respectively.
In terms of $\sigma _{i}$, Eq. (\ref{z.5.2}) reads%
\begin{equation}
A(k_{1},\cdots ,k_{n})=C\int_{0}^{\infty }\dif\tau _{2}(\tau
_{2})^{-d/2}\prod_{\mu =2}^{4}\int \frac{\dif\phi _{i}\dif\left\vert \sigma
_{i}\right\vert }{4\pi ^{2}\left\vert \sigma _{i}\right\vert }\exp \left(
-\sum_{i<j}k_{i}\cdot k_{j}\tilde{G}_{ij}\right) \text{ ,}  \label{z.5.8}
\end{equation}%
where%
\begin{equation}
\tilde{G}_{ij}=\frac{1}{2}\ln \left\vert \frac{\sigma _{i}}{\sigma _{j}}%
\right\vert ^{2}\frac{\sigma _{i}+\sigma _{j}}{\sigma _{i}-\sigma _{j}}-%
\frac{\pi }{2 \tau _{2}}\left( \frac{1}{2\pi \mi}\ln \left\vert \frac{\sigma _{i}%
}{\sigma _{j}}\right\vert ^{2}\right) ^{2} \text{ .} \label{z.5.9}
\end{equation}%
The integrations over $x_{i}$'s are equivalent to those over $\phi _{i}$, so we just need to consider the first term on the r.h.s. of Eq.(\ref{z.5.9}).
In the case (i), for any $i<j$, we have $y_{i}<y_{j}$ and hence $|\sigma _{i}|>|\sigma_{j}|$, thus we obtain the power series expression for the first term of $\tilde{G}_{ij}$:
\[
\frac{\sigma _{i}+\sigma _{j}}{\sigma _{i}-\sigma _{j}}=1+2\sum_{n=1}^{%
\infty }\left\vert \frac{\sigma _{j}}{\sigma _{i}}\right\vert
^{n}e^{\mi n(\phi _{j}-\phi _{i})}\text{ .}
\]%
Integrating out $\phi _{i}$'s we derive the reduced effective Green functions
\[
\tilde{G}_{ij}^{\text{eff}}=\frac{1}{2 }\ln \left\vert \frac{\sigma _{i}%
}{\sigma _{j}}\right\vert ^{2}-\frac{\pi }{2\tau _{2}}\left( \frac{1}{2\pi \mi}%
\ln \left\vert \frac{\sigma _{i}}{\sigma _{j}}\right\vert ^{2}\right) ^{2}%
\text{ .}
\]%
Note that $\tilde{G}_{ij}^{\text{eff}}$
are not symmetric with respect to indices $i$ and $j$ any longer - the order of indices is now important.

Now after integrating out the coordinates $x_{i}$'s in Eq. (\ref{z.5.2}), we get
\begin{align}
A(k_{1},\cdots ,k_{n})
&=C\int_{0}^{\infty }\dif\tau _{2}(\tau _{2})^{-d/2}  \nonumber \\
&\quad\times \prod_{\mu =2}^{4}\int_{0}^{\tau_2}\dif y_{\mu }\exp \left(
-\sum_{i<j}k_{i}\cdot k_{j}\left[y_{ij}+\frac{1}{\tau _{2}}(y_{ij})^{2}%
\right] 2\pi \right) \prod_{i}\theta(y_{i+1,i})\text{ ,}  \label{z.5.10}
\end{align}%
where step function $\theta(s)$ has the value $+1$ for $s>0$ and $0$ for $s<0$ and
 $y_{ij}\equiv y_{i,j}\equiv y_i-y_j$.
For comparison  with the Schwinger form in field theory, it is convenient to
make some variable substitutions. First, rescaling the $y_i$'s by a factor
of $\tau _{2}$, we get%
\begin{align}
A(k_{1},\cdots ,k_{n})=&C\int_{0}^{\infty }\dif\tau _{2}(\tau
_{2})^{-d/2+3}\nonumber\\
&\times\prod_{\mu =2}^{4}\int_{0}^{1}\dif y_{\mu }\exp \left(
-\sum_{i<j}k_{i}\cdot k_{j}\left[ y_{ij}+(y_{ij})^{2}\right] 2\pi \tau
_{2}\right)\prod_{i}\theta(y_{i+1,i}) \text{ .}
\end{align}%
Next, rescaling $\tau _{2}$ by a factor of $1/2\pi $, this gives an
extra factor $(2\pi )^{4-d/2}$ which can be absorbed into the overall constant
$C$. Then we obtain the final result of the amplitude%
\begin{align}
A(k_{1},\cdots ,k_{n})=&C\int_{0}^{\infty }\dif \tau _{2}(\tau
_{2})^{-d/2+n-1}\nonumber\\
&\times\prod_{\mu =2}^{n}\int_{0}^{1}\dif y_{\mu }\exp \left(
-\sum_{i<j}k_{i}\cdot k_{j}\left[ y_{ij}+(y_{ij})^{2}\right] \tau _{2}\right)\prod_{i}\theta(y_{i+1,i})
\text{ ,}  \label{z.5.11}
\end{align}%
with $n=4$. Up to an overall constant, Eqs. (\ref{z.5.4}) and (\ref{z.5.11}) are
equal after identifying the string variables $y_{ij}$ to the particle variables $\alpha _{i}$:
\begin{equation}
\alpha _{i}=y_{i+1}-y_{i},\quad i=1,\cdots,n,\quad \text{letting}\quad y_{n+1}=1\text{ ,}  \label{z.5.12}
\end{equation}
here we specify ourselves to the case $n=4$.
Using the identification (\ref{z.5.12}) in Eq. (\ref{z.5.11}), and eliminating $k_{4}$ by the conservation of momentum $\sum k_{i}=0$, $k_{2}\cdot k_{3}$ by the Mandelstam relation $k_{1}\cdot k_{2}+k_{1}\cdot k_{3}+k_{2}\cdot k_{3}=0$,
we obtain the expression for the amplitude corresponding to figure (a) in Fig.\ref{fig:2}%
\begin{align}
A(k_{1},\cdots ,k_{n}) &=C\int_{0}^{\infty }\dif\tau _{2}(\tau
_{2})^{-d/2+3}\prod_{i=1}^{4}\int_{0}^{1}\dif \alpha _{i} \,\delta\left(\sum_{i=1}^{4}\alpha_{i}-1\right)\nonumber \\
&\quad \exp \left\{ \left[ \left((\alpha _{1}+\alpha _{2})(\alpha _{2}+\alpha
_{3})-\alpha _{2}\right)2k_{1}\cdot k_{2}+2\alpha _{1}\alpha _{3}k_{1}\cdot k_{3}%
\right] \tau _{2}\right\}  \text{ .} \label{z.5.13}
\end{align}
It is clear now that if we specify the order of $y_{i}$'s along the integration interval, we can obtain the amplitude for the associated Feynman diagram. There are six different orders for $z_{i}$ since the coordinate  $z_{1}$ is fixed. These six different orders reduce to three Feynman diagrams up to the topological equivalence.

Based on the connection (\ref{z.5.12}), it is straightforward to generalize this result to the case where $n$ takes  any positive integers, proving the $n$-gons conjecture proposed in \cite{casali2014infrared}. Similar to the procedure above, we just need to show that the correspondence between case (i) and figure (a) still hold for generic value of $n$. On the field theory side, the generalization of (\ref{z.5.4}) for generic $n$ is
\begin{align}
A(k_{1},\cdots k_{n})
&=\int \frac{\dif^{d}p}{(2\pi )^{d}}\prod_{i=1}^{n}\int_{0}^{1}\dif \alpha _{i} \,\delta\left(\sum_{i=1}^{n}\alpha_{i}-1\right)\int_{0}^{\infty }\dif\alpha
\,\alpha ^{n-1} \nonumber \\
&\quad\exp \left( -\alpha \left( \sum_{i}^{n-1}\alpha _{i}(p+q_{i})^{2}+\left(
1-\sum_{i}^{n-1}\alpha _{i}\right) p^{2}\right) \right) \text{ ,} \label{z.5.14}
\end{align}
where
$
q_{i}=\sum_{j=1}^{i}k_{j}\text{ .}
$
After the substitution (\ref{z.5.12}) and the integration of $p$, this amplitude (\ref{z.5.14}) becomes
\begin{align}
A(k_{1},\cdots k_{n})
&=\prod_{\mu =2}^{n}\int_{0}^{1}\dif y_{\mu } \prod_{i}\theta(y_{i+1,i})\int_{0}^{\infty }\dif\alpha
\,\alpha ^{n-1-d/2} \nonumber \\
&\quad\exp \left(\alpha \left[(\sum_{i=1}^{n}k_{i}y_i)^2+\sum_{i<j}2 y_j k_i \cdot k_{j}\right]\right)   \text{ .} \label{z.5.15}
\end{align}
We arrive at Eq. (\ref{z.5.11}) after identifying $\alpha$ with $\tau_2$, so the two amplitudes are the same up to an overall constant.
This proves the n-gons conjecture.

\section{Conclusion and discussion}
We have tried to generalize the CHY formalism to the one loop level by means of the Green function method in null string theory.
We have found that the integrand is not modular invariant under the $S$ modular transformation. It is very important to check that the null string theory approach produces the same results as those obtained by field theory methods. Instead of deriving the off-shell scattering equations and solving them algebraically, we have chosen to perform the integration in the modular space explicitly, verifying the equivalence directly in the simplest $n$-gons case. Of course, in more general cases one has to calculate Feynman diagrams other than the ones for $n$-gons and the factor $\mathcal{I}(\tau,\{z_{i}\})$ in eq.(\ref{z.1.3}), being a function of the modular parameters, can no longer be taken as a constant.  Although the Green function approach still applies for the general cases, the calculation is expected to be more involved as one has to include world-sheet fermions and Kac-Moody currents in the treatment.  Being holomorphic, these fields  would not change the scattering equations but would contribute to the calculation of the other Feynman diagrams not related to the $n$-gons.

In a forthcoming paper \cite{yzz2017operator}, we will develop an operator formalism for the computation of one-loop amplitudes, based on the framework of the null string on a torus. There the BRST charge is constructed by considering the ghost contributions.

It would be interesting to generalize the null string formulation to higher genus Riemann surfaces which gives rise to multi-loop scattering amplitudes in point particle field theory. It is expected that the Green function method can be implemented with the help of the theta functions on the Riemann surfaces. Again, one would not expect a modular invariant integrand and thus the integration region could be simplified, in contrast to the ordinary string theory. Similar to the one-loop case, the physical meaning of the modular parameter would be as follows: the real part, the integration of which will project to the zero modes, is associated with the periodicity along the string direction, while the imaginary part is the Schwinger proper time in point particle field theory. For the scattering equation, one would expect more punctures with off-shell momenta on the Riemann sphere to be paired and sewed together. The situation is similar to the sewing in ordinary string theory but the calculation is anticipated to be much simpler, since only the zero modes survive propagating in the pinching limit.

\acknowledgments
We are grateful to Piotr Tourkine, Song He and Yihong Gao for useful discussions on the relevant subjects. This work is supported in part by the National Natural Science Foundation of China Grants No.11675240 and partly supported by Key Research Program of Frontier Sciences, CAS. YZZ would like to thank the Institute of Theoretical Physics, Chinese Academy of Sciences, for hospitality and support. He also acknowledges the partial support of the Australian Research Council Discovery Project DP140101492.

\appendix
\section{Green functions}
In section 2, the Green function of the operator $\bar{\partial}^{2}$ on the cylinder is derived using the operator formalism. In this appendix, we shall generalize the result to the torus  boundary conditions.
In the limit $z\to 0$,%
\begin{equation*}
G(z)=\frac{\bar{z}}{z}
\end{equation*}%
obviously satisfies the Green function's equation
\[
\bar{\partial}^{2}\left(\frac{\bar{z}}{z}\right)= 2\pi\delta^{2}(z)\text{ .}
\]
This solution can be generalized to the infinite cylinder by imposing the periodicity $z\to z+1$. Hence, a natural construction of Green function on cylinder is
\[
G(z) =\sum_{n\in\mathbb{Z}}\frac{\bar{z}+n}{z+n}
=\bar{z}\sum_{n\in\mathbb{Z}}\frac{1}{z+n}+\sum_{n\in\mathbb{Z}}\frac{n}{z+n}+\text{regularizations} \text{ .}
\]
By the use of of
\[
\sum_{n\in\mathbb{Z}}\frac{1}{z+n}=\pi \cot \pi z
\]%
the first term becomes $\bar{z}\pi \cot (\pi z)$. The second term is
still divergent. However, if we subtract the infinity by adding $-1$ to the second summand term, it can be
regularized to a well-defined function which still is the solution of Eq. (\ref{z.2.1}):
\[
\sum_{n}\frac{n}{z+n} \rightarrow \sum_{n}\left( \frac{n}{z+n}-1\right)
=\sum_{n}\frac{-z}{z+n} =-z\pi \cot \pi z
\]
so we get the solution (\ref{z.0.5})%
\[
G(z)=(\bar{z}-z)\pi \cot \pi z \text{ .}
\]%

We can use the same trick for torus, the solution is%
\begin{align*}
G(z) &=\sum_{n,m}\left(\frac{\bar{z}+n+m\bar{\tau}}{z+n+m\tau }-1 \right)
=\sum_{n,m}\frac{\left( \bar{z}-z\right) }{z+n+m\tau }+\sum_{n,m}\frac{m(%
\bar{\tau}-\tau )}{z+n+m\tau }  \nonumber \\
&=(\bar{z}-z)\sum_{n,m}\frac{1}{z+n+m\tau }+(\bar{\tau}-\tau )\partial
_{\tau }\left( \sum_{n,m}\ln (z+n+m\tau )\right)   \nonumber \\
&\sim (\bar{z}-z)\frac{\theta _{1}^{\prime }(z|\tau )}{\theta _{1}(z|\tau )}%
+(\bar{\tau}-\tau )\partial _{\tau }\ln \theta _{1}(z|\tau )  \nonumber \\
&=(\bar{z}-z)\partial _{z}\ln \theta _{1}(z|\tau )+(\bar{\tau}-\tau
)\partial _{\tau }\ln \theta _{1}(z|\tau )
\end{align*}%
where we have used%
\begin{equation*}
\sum_{n,m}\frac{1}{z+n+m\tau }\sim \frac{\theta _{1}^{\prime }(z|\tau )}{%
\theta _{1}(z|\tau )}  \label{5.12}
\end{equation*}%
The symbol `$\sim$' means these two functions share the same pole structure. The quadratic term of Green function can be easily obtained by the double periodicity.

\newcommand*{\doi}[1]{\href{http://dx.doi.org/#1}}
\providecommand{\href}[2]{#2}\begingroup\raggedright\endgroup

\end{document}